\documentclass[letterpaper, 10 pt, conference]{ieeeconf}  % Comment this line out
\IEEEoverridecommandlockouts                              % This command is only
\usepackage{epsfig} % for postscript graphics files
\usepackage{times} % assumes new font selection scheme installed
\usepackage{amsmath} % assumes amsmath package installed
\usepackage{amssymb}  % assumes amsmath package installed
\usepackage{mathtools}
\usepackage{ntheorem}
\usepackage{arydshln}

\newtheorem{definition}{Definition}

\newtheorem{lemma}{Lemma}

\DeclareMathOperator*{\diag}{diag}

\usepackage{soul}

\title{\LARGE \bf
Impact Sensitivity Analysis of Cooperative Adaptive Cruise Control Against Resource-Limited Adversaries}

\author{Mischa Huisman, Carlos Murguia, Erjen Lefeber, and Nathan van de Wouw% <-this % stops a space
\thanks{The research leading to these results has received funding from the European Union’s Horizon Europe programme under grant agreement No 101069748 – SELFY project.}% <-this % stops a space
\thanks{M. Huisman, C. Murguia, E. Lefeber, and N. van de Wouw are with the Department of Mechanical Engineering, Eindhoven University of Technology, The Netherlands.
        {\tt\small [m.r.huisman}, {\tt\small C.G.Murguia}, {\tt\small A.A.J.Lefeber}, {\tt\small N.v.d.Wouw]@tue.nl}}
}

\begin{document}

\maketitle
\thispagestyle{empty}
\pagestyle{empty}

%%%%%%%%%%%%%%%%%%%%%%%%%%%%%%%%%%%%%%%%%%%%%%%%%%%%%%%%%%%%%%%%%%%%%%%%%%%%%%%%
\begin{abstract}
Cooperative Adaptive Cruise Control (CACC) is a technology that allows groups of vehicles to form in automated, tightly-coupled platoons. CACC schemes exploit Vehicle-to-Vehicle (V2V) wireless communications to exchange information between vehicles. However, the use of communication networks brings security concerns as it exposes network access points that the adversary can exploit to disrupt the vehicles' operation and even cause crashes. In this manuscript, we present a sensitivity analysis of CACC schemes against a class of resource-limited attacks. We present a modelling framework that allows us to systematically compute outer ellipsoidal approximations of reachable sets induced by attacks. We use the size of these sets as a security metric to quantify the potential damage of attacks affecting different signals in a CACC-controlled vehicle and study how two key system parameters change this metric. We carry out a sensitivity analysis for two different controller implementations (as given the available sensors there is an infinite number of realizations of the same controller) and show how different controller realizations can significantly affect the impact of attacks. We present extensive simulation experiments to illustrate the results.
\end{abstract}

%%%%%%%%%%%%%%%%%%%%%%%%%%%%%%%%%%%%%%%%%%%%%%%%%%%%%%%%%%%%%%%%%%%%%%%%%%%%%%%%
\section{INTRODUCTION}
%% 8 page version
%\input{V2P8/1.IntroductionP8}

%% 6 page version
Connected and Automated Vehicles (CAVs) have become a promising technology in the automotive industry, offering potential improvements in safety, mobility, and environmental sustainability. Cooperative Adaptive Cruise Control (CACC) is a well-explored technology within CAVs that allows groups of vehicles to form tightly-coupled platoons by exchanging inter-vehicle data through Vehicle-to-Vehicle (V2V) wireless communication networks \cite{Ploeg2014a}. However, the use of communication networks brings security concerns as it exposes network access points that the adversary can exploit via cyberattacks, resulting in new safety and security challenges that were not encountered in traditional vehicle systems \cite{Amoozadeh2015}-\cite{sun_survey_2022}. Therefore, a pressing need arises for technologies that can quantify the potential impact of cyberattacks on platooning behavior. Moreover, quantify the impact sensitivity against the potentially compromised elements, e.g., sensors, actuators, networks, and software, and provide guidelines to allocate security resources to minimize the impact of attacks.

New technologies in this field focus on the prevention and detection of cyberattacks, which are well-explored research fields, where different sorts of attacks and methods are applied to increase the security of the system \cite{sun_survey_2022}-\cite{ju_survey_2022}. However, prevention and detection methods are bounded by unknown process and measurement disturbances, leaving space for the adversary to perform resource-limited stealth attacks \cite{zhang_networked_2019}. 

There is limited research regarding the potential impact such a resource-limited attack has on platooning behavior. In \cite{Amoozadeh2015} and \cite{vdHeijden2017}, the authors present simulation studies where they compare the impact of particular types of attacks on cooperative driving systems. Although interesting, it lacks generality in terms of the variety of attacks that can be considered. To overcome these limitations, reachable sets induced by general resource-limited attacks (referred hereto as adversarial reachable sets) can be used to analyze attackers' capabilities to drive the platoon to unsafe states. In \cite{Dadras2018}, adversarial reachable sets are approximated in a simulation environment for a class of bounded attacks. In \cite{Murguia2020} and \cite{kafash_constraining_2018}, for a general class of linear-time invariant dynamical systems driven by peak-bounded disturbances, ellipsoidal outer approximations of the reachable sets are obtained using first-principles models and convex optimization techniques. 

In this manuscript, we develop a framework to analyze the impact of cyberattacks by a general class of resource-limited adversaries on standard CACC schemes, using the outer ellipsoidal approximation of the adversarial reachable set. Using the size of these sets as a security metric, we quantify the potential damage of attacks affecting different signals in a CACC-controlled vehicle. In this scope, we investigate how the use of different sensors and key system parameters affects the reachable set, where a difference in size can indicate a more resilient implementation. The notion of critical states is introduced, which characterize states that, if reached, compromise the safety and integrity of the vehicle. To support our claims, we have conducted an extensive simulation study on two different realized CACC controllers. Our findings reveal that the impact of attacks can significantly vary depending on the type of controller implementation used. 

The structure of the paper is as follows: Section \ref{sec:Preliminaries} introduces some preliminary results necessary for the subsequent sections. Section \ref{sec:SystemDescription} describes the platooning model, the CACC scheme, and the available measurements. Section \ref{sec:AdversarialLTI} presents the adversarial LTI systems derived from the system description. In Section \ref{sec:SecurityAssessment}, an extensive simulation study is conducted to demonstrate how different implementations and system parameters can significantly affect the impact of attacks. Finally, Section \ref{sec:Conclusion} provides the concluding remarks.

%% UNUSED REFS
%CACC schemes are able to achieve string stability \cite{Ploeg2014}, \cite{Ploeg2014a}, enhance vehicle throughput \cite{Santhanakrishnan2003}, and reduce fuel consumption \cite{Alam2010}.

% The distinct characteristics of cyberattacks require different approaches to enhance the system's resiliency. Increasing resilience can be accomplished primarily through three actions: prevention, detection, and mitigation \cite{teixeira_secure_2015}. Both prevention and detection are well-explored research fields, where different sorts of attacks and methods are applied to increase the resilience of the system \cite{sun_survey_2022}, \cite{chowdhury_attacks_2020}, \cite{ju_survey_2022}. However, prevention and detection methods are bounded by unknown process and measurement disturbances, leaving space for the adversary to perform stealth attacks \cite{zhang_networked_2019}. Mitigation of cyberattacks on the CAVs control scheme can bridge the gap where prevention and detection methods fail, however, there is limited research in this field.

%%%%%%%%%%%%%%%%%%%%%%%%%%%%%%%%%%%%%%%%%%%%%%%%%%%%%%%%%%%%%%%%%%%%%%%%%%%%%%%%
\section{Mathematical Preliminaries} \label{sec:Preliminaries}
%% 8 page version
%\input{V2P8/2.PreliminariesP8}

%% 6 page version
\subsection{Notation}
The symbol $\mathbb{R}$ stands for the real numbers, $\mathbb{R}_{>0}$($\mathbb{R}_{\geq 0}$) denotes the set of positive (non-negative) real numbers. The symbol $\mathbb{N}$ stands for the set of natural numbers, including zero. The $n \times m$ matrix composed of only zeros is denoted by $\mathbf{0}_{n \times m}$, or simply $\mathbf{0}$ when its dimension is clear. Consider a finite index set $\mathcal{L} := \{l_1,\ldots,l_\rho\}  \subset \mathbb{N}$ with cardinality $\text{card}[\mathcal{L}] = \rho$, e.g., $\mathcal{L} = \{1,3,7,15\}$ with $\text{card}[\mathcal{L}] = 4$, the notation diag[$B_j$] and $(B_j)$, $j \in \mathcal{L}$, stand for the diagonal block matrix $\text{diag}[B_{l_1},\ldots,B_{l_{\rho}}]$ and stacked block matrix $(B_{l_1},\ldots,B_{l_{\rho}})$, respectively. The notation $A \geq 0$ (resp., $A \leq 0$) indicates that the matrix $A$ is positive (resp., negative) semidefinite, i.e., all the eigenvalues of the symmetric matrix $A$ are positive (resp. negative) or equal to zero, whereas the notation \(A > 0\) (resp., \(A < 0\)) indicates the positive (resp., negative) definiteness, i.e., all the eigenvalues are strictly positive (resp. negative). We often omit implicit time dependencies of signals for simplicity of notation.

\subsection{Definitions and Preliminary Results}
%%% DEFENITION 1 (REACHABLE SET) %%%
\begin{definition}[Reachable Set]\emph{\cite{Murguia2020}}
Consider the perturbed Linear Time-Invariant (LTI) system:\label{def1}
\begin{equation}
    \label{eq:LTI_set}
    \zeta(k+1) = \mathcal{A} \zeta(k) + \sum_{i=1}^N \mathcal{B}_i \omega_i(k),
\end{equation}
with $k\in\mathbb{N}$, state $\zeta(k) \in \mathbb{R}^{n_\zeta}$, perturbation $\omega_i \in \mathbb{R}^{p_i}$ satisfying $\omega_i^{\top} W_i \omega_i \leq 1$ for some positive definite matrix $W_i \in \mathbb{R}^{p_i \times p_i}, i = \{1,...,N\}, N\in \mathbb{N}$, and matrices $\mathcal{A} \in \mathbb{R}^{n_\zeta \times n_\zeta}$ and $\mathcal{B}_i \in \mathbb{R}^{n_\zeta \times p_i}$. The reachable set $\mathcal{R}^\zeta$(k) at time $k\geq0$ from the initial condition $\zeta_0 \in \mathbb{R}^{n_\zeta}$ is the set of states reachable in $k$ steps by system \eqref{eq:LTI_set} through all possible disturbances satisfying $\omega_i^{\top} W_i \omega_i \leq 1$, i.e.,
\begin{equation}
    \mathcal{R}^{\zeta}(k) \! \coloneqq \!\!\left\{ \zeta \in \mathbb{R}^{n_\zeta} \left| \begin{array}{l}
        \zeta=\zeta(k), \zeta(k) \text{ \textit{solution to \eqref{eq:LTI_set},}}\\
         \text { \textit{and} } \omega_i(k)^{\top} W_i \omega_i(k) \leq 1,
    \end{array} \!\!\!\!\!  \right\} \right. .
\end{equation}
\end{definition}
%%% END DEFENITION 1 (REACHABLE SET) %%%

%%% LEMMA 1 (Minimum Volume Ellipsoidal Approximation %%%
\vspace{2mm}
\begin{lemma}[Ellipsoidal Approximation]\emph{\cite{Murguia2020}}\label{lemma1}
	Consider the perturbed LTI system \eqref{eq:LTI_set} and the reachable set $\mathcal{R}^{\zeta}(k)$ in Definition \ref{def1}. For a given $a\in(0,1)$, if there exist constants $a_{1}$, $\ldots$, $a_{N}$ and matrix $P$ that is the solution of the convex program:
	\begin{equation}\label{lmi1}
		\left\{\!\begin{aligned}
            &\min_{P,a_{1},\ldots, a_{N}}-\log\det[P],\\
			&\text{\emph{s.t.}} \hspace{1mm}a_{1},\ldots, a_{N}\in(0,1), \hspace{.5mm} a_{1}+\ldots+ a_{N}\geq a,\\
			&P>0, \begin{bmatrix}
				aP&\mathcal{A}^{\top}P&\mathbf{0}\\
				P\mathcal{A}&P&P\mathcal{B}\\
				\mathbf{0}&\mathcal{B}^{\top}P&W_{a}
			\end{bmatrix}\geq 0,
		\end{aligned}\right.
	\end{equation}
	with matrices $W_{a}:=\diag\begin{bmatrix}
		(1-a_{1})W_{1},\ldots,(1-a_{N})W_{N}
\end{bmatrix}\in\mathbb{R}^{\bar{p}\times\bar{p}}$ and $\mathcal{B}:=(\mathcal{B}_{1},\ldots,\mathcal{B}_{N})\in\mathbb{R}^{n_{\zeta}\times \bar{p}}$, and $\bar{p}=\sum_{i=1}^{N}p_{i}$; then, for all $k \in \mathbb{N}$, $\mathcal{R}^{\zeta}(k)\subseteq \mathcal{E}^{\zeta}(k)$ with $\mathcal{E}^{\zeta}(k):=\{ \zeta^{\top}(k)P^{\zeta}\zeta(k) \leq\alpha^{\zeta}(k)\} $, with convergent scalar sequence $\alpha^{\zeta}(k):=a^{k-1}\zeta(k)^{\top}P\zeta(k)+((N-a)(1-a^{k-1}))/(1-a)$. Ellipsoid $\mathcal{E}^{\zeta}(k)$ has the minimum asymptotic volume among all outer ellipsoidal approximations of $\mathcal{R}^{\zeta}(k)$.
\end{lemma}
%%% END LEMMA 1 %%%

%%% LEMMA 2 PROJECTION %%%
\vspace{2mm}
\begin{lemma}[Projection]\emph{\cite{Murguia2020}}\label{lemma2}
	Consider the ellipsoid:
	\begin{equation}
		\begin{split}
			\mathcal{E}:=\left\lbrace x\in\mathbb{R}^{n}, y\in\mathbb{R}^{m}\bigg|\begin{bmatrix}
				x\\y
			\end{bmatrix}^{\top}\begin{bmatrix}
				Q_{1}& \hspace{-1mm}Q_{2}\\Q_{2}^{\top}&\hspace{-1mm}Q_{3}
			\end{bmatrix}\begin{bmatrix}
				x\\y
			\end{bmatrix}=\alpha \right\rbrace ,
		\end{split}
	\end{equation}
	for some positive definite matrix $Q\in\mathbb{R}^{(n+m)\times(n+m)}$ and constant $\alpha\in\mathbb{R}_{>0}$. The projection $\mathcal{E}'$ of $\mathcal{E}$ onto the $x$-hyperplane is given by the ellipsoid:
	\begin{equation}
		\mathcal{E}':=\left\lbrace x\in \mathbb{R}^{n} \left| \hspace{1mm} x^{\top}\begin{bmatrix}
			Q_{1}-Q_{2}Q_{3}^{-1}Q_{2}^{\top} 
		\end{bmatrix}x=\alpha\right\rbrace . \right.
	\end{equation}
\end{lemma}
\vspace{2mm}
%%% END LEMMA 2%%%

%%%%%%%%%%%%%%%%%%%%%%%%%%%%%%%%%%%%%%%%%%%%%%%%%%%%%%%%%%%%%%%%%%%%%%%%%%%%%%%%
\section{System Description}\label{sec:SystemDescription}
%% 8 page version
%\input{V2P8/3.ModelDescriptionP8}

\begin{figure}[bt]\centering
		\includegraphics[width=\linewidth]{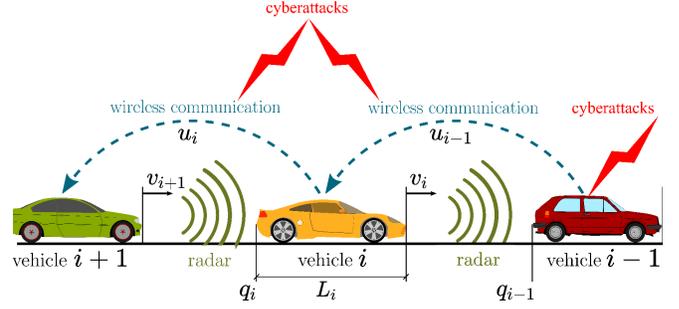}
		\caption{CACC-equipped vehicle platoon. Each vehicle is equipped with onboard sensors (e.g., radars/LiDARs, cameras, and velocity/acceleration sensors). Vehicles may be subject to FDI attacks.}
		\centering
		\label{fig:Platoon}
  \vspace{-6mm}
\end{figure}

%% 6 page version
Consider a platoon of $m$ vehicles, schematically depicted in Fig. \ref{fig:Platoon}, with $d_i(t) = q_{i-1} - q_i - L_i$ ($q_i$ reflects the position of the rear bumper of vehicle $i$ and $L_i$ its length) being the distance between vehicle $i$ and its preceding vehicle $i-1$, $v_i$ the velocity of vehicle $i$. The objective of each vehicle is to keep a desired distance $d_{r,i}$ (the so-called spacing policy) with its preceding vehicle:
\begin{align}
    d_{r,i}(t) = r_i + h v_i(t), \, i \in S_m,
\end{align}
with the headway $h>0$, standstill distance $r_i>0$, and $S_m \coloneqq \{ i \in \mathbb{N} \mid 1 \leq i \leq m \}$ (i.e., the set of all vehicles in a platoon of length $m \in \mathbb{N}$). A set of homogeneous vehicles is assumed; therefore, also $h$ is chosen the same for all $i$. The spacing error is then defined as 
\begin{align}
\label{eq:ErrorFormulation}
    \begin{split}
        e_i(t)  &\coloneqq d_i(t) - d_{r,i}(t).  
    \end{split}
\end{align}

We adopt the CACC scheme introduced in \cite{Ploeg2014a}, which considers a longitudinal vehicle model where the dynamics are described as
\begin{align}
\label{eq:LongVehModel}
    \begin{bmatrix}
        \dot{q}_i \\ \dot{v}_i \\\dot{a}_i
    \end{bmatrix}
    = 
    \begin{bmatrix}
        v_i(t) \\ a_i(t) \\ -\frac{1}{\tau}a_i(t) + \frac{1}{\tau} u_i(t) 
    \end{bmatrix}, \, i \in S_m,
\end{align}
where $\tau$ is a time constant modelling driveline dynamics, $a_i(t)$ denotes the acceleration of vehicle $i$, and $u_i(t)$ is its desired acceleration (the control input). The controller in \cite{Ploeg2014a} is a dynamic controller of the following form:
\begin{equation}
\label{eq:Controller0}
    \mathcal{C} \coloneqq \dot{u}_i = -\tfrac{1}{h} u_i + \tfrac{k_p}{h} e_i + \tfrac{k_d}{h} \dot{e}_i + \tfrac{k_{dd}}{h} \ddot{e}_i + \tfrac{1}{h}u_{i-1}, 
\end{equation}
where $k_p, k_d$, and $k_{dd}$ are the controller gains. We refer to $\mathcal{C}$ as the base controller.

The real-time realization of any control scheme depends on the available sensors $y_{i,j}$ (sensor number $j$ of vehicle $i$). We use a combination of sensor data coming from onboard sensors (e.g., radar, LiDAR, cameras, and velocity/acceleration sensors) and received data transmitted wirelessly between adjacent vehicles. We assume that the sensors available to implement control actions are:
\begin{subequations}
\label{eq:SensorMeasurements}
    \begin{align}
        y_{i,1} &\coloneqq q_{i-1}(t) - q_{i}(t) - L_i + \delta_{i,1}(t), \\
        y_{i,2} &\coloneqq v_i(t) + \delta_{i,2}(t), \\
        y_{i,3} &\coloneqq a_i(t) + \delta_{i,3}(t), \\
        y_{i,4} &\coloneqq v_{i-1}(t) - v_i(t) + \delta_{i,4}(t), \\
        y_{i,5} &\coloneqq a_{i-1}(t) + \delta_{i,5}(t), \\
        y_{i,6} &\coloneqq u_{i-1}(t) + \delta_{i,6}(t).
    \end{align}
\end{subequations}
Herein, $\delta_{i,j}(t)$ models potential false data injection attacks and $j \in \{1, ..., 6\}$. It is important to note that sensors $y_{i,1}$ and $y_{i,4}$ provide relative tracking and relative velocity information, $y_{i,2}$ and $y_{i,3}$ are the onboard measured velocity and acceleration, and sensors $y_{i,5}$ and $y_{i,6}$ model data received wirelessly from the preceding vehicle via V2V communication. Using the measured values in \eqref{eq:SensorMeasurements}, base controller \eqref{eq:Controller0} can be implemented as follows:
\begin{align}\label{eq:ui1}
&\mathcal{C}_1 \coloneqq \left\{ 
    \begin{aligned}
        \dot{\xi}_{i,1} &= -(\tfrac{1}{h} + \tfrac{k_{dd}}{\tau})\xi_{i,1} + \tfrac{k_p}{h} y_{i,1} - k_p y_{i,2} \\ 
                        & \,\,\,\,\,\,\, - (k_d + \tfrac{k_{dd}}{h} - \tfrac{k_{dd}}{\tau}) y_{i,3} + \tfrac{k_d}{h}y_{i,4} \\
                        & \,\,\,\,\,\,\, + \tfrac{k_{dd}}{\tau}y_{i,5} + \tfrac{1}{h} y_{i,6} - \tfrac{k_p}{h} r_i \\
        {u}_{i} &= {\xi}_{i,1},
    \end{aligned} \right.
\end{align}
with controller state $\xi_{i,1} \in \mathbb{R}$. 

We remark that the controller implementation in \eqref{eq:ui1} is not unique, and refer to such a controller as a realized controller. An infinite number of equivalent realizations of the base controller $\mathcal{C}$ exist that result in the same control signal $u_i(t)$ at the vehicle for $\delta_{i,j} = 0$. For instance, consider the CACC controller in \cite{Lefeber2020}, with a change of coordinates $\tfrac{\tau}{h}\xi_{i,2} = \tfrac{\tau}{h}a_{i-1} + (1-\tfrac{\tau}{h})a_i - \xi_{i,1}$ and new controller state $\xi_{i,2} \in \mathbb{R}$. Applying this coordinate transformation to \eqref{eq:Controller0}, and using the available sensors in \eqref{eq:SensorMeasurements}, a second realized controller, $\mathcal{C}_2$, can be obtained as
\begin{align}\label{eq:ui2}
&\mathcal{C}_2 \coloneqq \left\{ 
    \begin{aligned}
        \dot{\xi}_{i,2} &= -\tfrac{1+k_{dd}}{\tau}\xi_{i,2} - \tfrac{k_p}{\tau} y_{i,1} + \tfrac{k_p h}{\tau} y_{i,2} \\ 
                        & \,\,\,\,\,\,\, + \tfrac{k_{d} h}{\tau} y_{i,3} - \tfrac{k_d}{\tau}y_{i,4} + \tfrac{k_p}{\tau} r_i, \\
        {u}_{i} &= \tfrac{\tau}{h}y_{i,5} + (1 - \tfrac{\tau}{h})y_{i,3} - \tfrac{\tau}{h}\xi_{i,2}.
    \end{aligned} \right.
\end{align}
That is, we have two different realizations, $\mathcal{C}_1$ and $\mathcal{C}_2$, of the CACC controller in \eqref{eq:Controller0} --- each result in the same control signal $u_i(t)$ applied to the vehicle for $\delta_{i,j}(t) = 0$. Note that $\mathcal{C}_2$ requires the actual acceleration of the preceding vehicle, whereas $\mathcal{C}_1$ requires the control input of the preceding vehicle. However, when $\delta_{i,j}(t) \neq 0$, for some $t\geq 0$, they will, in general, lead to different control signals $u_i(t)$.

To quantify the impact of the attack on the two different controller realizations, we are interested in the reachable sets induced by the attack signals $\delta_{i,j}(t)$ and use the size of these sets as a security metric. Here, we are interested in assessing whether $\mathcal{C}_1$ and $\mathcal{C}_2$ lead to different sensitivity to attacks for the same class of resource-limited adversaries for varying system and control parameters.

%Both $\mathcal{C}_1$ and $\mathcal{C}_2$ are evaluated by computing the adversarial reachable sets for varying system and control parameters, where a difference in size indicates a more resilient implementation. 

%\begin{equation}
%\label{eq:Controller0}
%    \mathcal{C} \coloneqq \left\{
%    \!\begin{aligned}%[b]
%        \dot{\eta}_i &= -\tfrac{1}{h} \eta_i + \tfrac{k_p}{h} e_i + \tfrac{k_d}{h} \dot{e}_i + \tfrac{k_{dd}}{h} \ddot{e}_i + \tfrac{1}{h}u_{i-1}, \\
%        u_i &= \eta_i,
%    \end{aligned} \right.
%\end{equation}

%%%%%%%%%%%%%%%%%%%%%%%%%%%%%%%%%%%%%%%%%%%%%%%%%%%%%%%%%%%%%%%%%%%%%%%%%%%%%%%%
\section{Adversarial LTI System}\label{sec:AdversarialLTI}
%% 8 page version
%\input{V2P8/4.PerturbedLTIP8}

%% 6 page version
\subsection{Closed-Loop Platooning Dynamics}

Using the above expressions for the vehicle dynamics \eqref{eq:LongVehModel}, sensors \eqref{eq:SensorMeasurements}, and realized controllers $\mathcal{C}_1$ or $\mathcal{C}_2$, we can write the closed-loop tracking dynamics in terms of the attack signals $\delta_{i,j}(t)$ introduced in \eqref{eq:SensorMeasurements}. To make a fair comparison between the realized controllers $\mathcal{C}_1$ and $\mathcal{C}_2$, we write them in the same coordinates. First, consider \eqref{eq:ui1} and the change of coordinates $\xi_{i,1} = \tfrac{\tau}{h}a_{i-1} + (1-\tfrac{\tau}{h})a_i - \tfrac{\tau}{h}\zeta_{i,1}$ (this is the same transformation, though inverse, that relates $\mathcal{C}_1$ and $\mathcal{C}_2$ above), with new controller state $\zeta_{i,1} \in \mathbb{R}$. Then, after some computations, \eqref{eq:ui1} can be written in terms of $\zeta_{i,1}$ and $\delta_{i,j}$ as follows:
\begin{align}\label{eq:ui1_att}
&\mathcal{C}_1 \coloneqq \left\{ 
    \begin{aligned}
        %\dot{\zeta}_{i,1} &= -\tfrac{1+k_{dd}}{\tau} \zeta_{i,1} - \tfrac{1}{\tau}(k_p e_i + k_d \dot{e}_i) - \tfrac{k_p}{\tau} \delta_{i,1} \\
        \dot{\zeta}_{i,1} &= -\tfrac{1+k_{dd}}{\tau} \zeta_{i,1} - \tfrac{k_p}{\tau}e_i - \tfrac{k_d}{\tau}\dot{e}_i- \tfrac{k_p}{\tau} \delta_{i,1} \\
        & \,\,\,\,\,\,\,  + \tfrac{k_p h}{\tau} \delta_{i,2} + (\tfrac{k_d h + k_{dd}}{\tau} - \tfrac{k_{dd} h}{\tau^2})\delta_{i,3} \\
        & \,\,\,\,\,\,\, - \tfrac{k_d}{\tau}\delta_{i,4} - \tfrac{k_{dd}}{\tau}\delta_{i,5} - \tfrac{1}{\tau}\delta_{i,6}, \\
        {u}_{i} &= -\tfrac{\tau}{h}{\zeta}_{i,1} + \tfrac{\tau}{h}a_{i-1} + (1-\tfrac{\tau}{h}) a_i.
    \end{aligned} \right.
\end{align}

Define the stacked state vector $\hat{x}_{i} \coloneqq \text{col}[e_i, \dot{e}_i, \zeta_{i,1}, z_i]$, where $z_i(t) \coloneqq q_{i-1} - q_i - L_i - r_i$. Then, the resulting closed-loop dynamics \eqref{eq:LongVehModel}, \eqref{eq:SensorMeasurements}, \eqref{eq:ui2} can be written as follows:
\begin{equation}
\label{eq:LTI_C1_Cont}
    \dot{\hat{x}}_{i} \coloneqq A^c \hat{x}_{i} + B_v^c v_{i-1} + \sum_{j\in \mathcal{L}} \hat{\Gamma}_j^c \delta_{i,j}
\end{equation}
with system matrices
\begin{equation} \label{eq:SystemMatricesCL}
    \begin{split}
        A^c&:=\begin{bmatrix*}[r]
        0&1&0&0\\
        0&0&1&0\\
        -\frac{k_p}{\tau}&-\frac{k_d}{\tau}&-\frac{1+k_{dd}}{\tau}&0\\
        \frac{1}{h}&0&0&-\frac{1}{h}
        \end{bmatrix*},
        B_{v}^c:=\begin{bmatrix}
        0\\
        0\\
        0\\
        1
        \end{bmatrix},
    \end{split}
\end{equation}
and attack matrices $\hat{\Gamma}^c_j$
\begin{equation}
\label{eq:Attack_Gamma_C1}
	\left\{ \!\! 
 \begin{aligned}%[b]
 \hat{\Gamma}_{1}^c&\!:=\!\!\begin{bmatrix*}
			0\\
			0\\
			-\frac{k_p}{\tau}\\
			0
		\end{bmatrix*}\!\!\!, 
\hat{\Gamma}_{2}^c\!:=\!\!\begin{bmatrix*}
			0\\
			0\\
			\frac{k_ph}{\tau}\\
			0
		\end{bmatrix*}\!\!\!,  
\hat{\Gamma}_{3}^c\!:=\!\!\begin{bmatrix*}
			0\\
			0\\
			\frac{k_d h + k_{dd}}{\tau} - \frac{k_{dd} h}{\tau^2}\\
			0
		\end{bmatrix*}\!\!\!,  \\
 \hat{\Gamma}_{4}^c\!&:=\!\!\begin{bmatrix*}
			0\\
			0\\
			-\frac{k_d}{\tau}\\
			0
		\end{bmatrix*}\!\!\!, 
\hat{\Gamma}_{5}^c\!:=\!\!\begin{bmatrix*}
			0\\
			0\\
			-\frac{k_{dd}}{\tau}\\
			0
		\end{bmatrix*}\!\!\!, 
\hat{\Gamma}_{6}^c\!:=\!\!\begin{bmatrix*}
			0\\
			0\\
			-\frac{1}{\tau}\\
			0
		\end{bmatrix*}\!\!\!.
	 \end{aligned} \right.
\end{equation}
Here, we have introduced the index set $\mathcal{L} \subseteq \{1, ..., 6\}$, which denotes the set of compromised sensors in \eqref{eq:SensorMeasurements} (i.e. $\delta_{i,j}(t) \neq 0$ for $j \in \mathcal{L}$ and some $t\geq 0$). Similarly, $\mathcal{C}_2$ can be written as follows: 
\begin{align}\label{eq:ui2_att}
&\mathcal{C}_2 \coloneqq \left\{ 
    \begin{aligned}
        \dot{\xi}_{i,2} &= -\tfrac{1+k_{dd}}{\tau}\xi_{i,2} - \tfrac{k_p}{\tau} e_i - \tfrac{k_d}{\tau}\dot{e}_i - \tfrac{k_p}{\tau}\delta_{i,1} \\
        & \,\,\,\,\,\,\, + \tfrac{k_p h}{\tau}\delta_{i,2} + \tfrac{k_d h}{\tau}\delta_{i,3} - \tfrac{k_d}{\tau} \delta_{i,4}, \\
        {u}_{i} &= -\tfrac{\tau}{h}\xi_{i,2} + (1 - \tfrac{\tau}{h})a_i + \tfrac{\tau}{h}a_{i-1} \\
        & \,\,\,\,\,\,\, + (1-\tfrac{\tau}{h})\delta_{i,3} + \tfrac{\tau}{h}\delta_{i,5}.
    \end{aligned} \right.
\end{align}
Then, the closed-loop dynamics \eqref{eq:LongVehModel}, \eqref{eq:SensorMeasurements}, \eqref{eq:ui2_att} is given by
\begin{equation}
\label{eq:LTI_C2_Cont}
    \dot{\bar{x}}_{i} \coloneqq A^c \bar{x}_{i} + B_v^c v_{i-1} + \sum_{j\in \mathcal{L}} \bar{\Gamma}_j^c \delta_{i,j},
\end{equation}
where $\bar{x}_i \coloneqq \text{col} [e_i,\dot{e}_i,\xi_{i,2},z_i]$, $\bar{\Gamma}_1^c = \hat{\Gamma}_1^c$, $\bar{\Gamma}_2^c = \hat{\Gamma}_2^c$, $\bar{\Gamma}_4^c = \hat{\Gamma}_4^c$, $A^c$ and $B_v^c$ as defined in \eqref{eq:SystemMatricesCL}, and
\begin{equation}
\label{eq:Attack_Gamma_C2}
	\!\begin{aligned}%[b]
\bar{\Gamma}_{3}^c:=\begin{bmatrix*}
			0\\
			1-\frac{h}{\tau}\\
			\frac{k_dh}{\tau}\\
			0
		\end{bmatrix*}, 
\bar{\Gamma}_{5}^c:=\begin{bmatrix*}
			0\\
			-1\\
			0\\
			0
		\end{bmatrix*}, 
\bar{\Gamma}_{6}^c:=\begin{bmatrix*}
			0\\
			0\\
			0\\
			0
		\end{bmatrix*}.
	\end{aligned} 
\end{equation}

Because platooning controllers run in discrete time, and attacks operate on sampled signals, discrete-time equivalent models of \eqref{eq:LTI_C1_Cont} and \eqref{eq:LTI_C2_Cont} are obtained via exact discretization at the sampling time instant, $t = T_s k, \, k \in \mathbb{N}$, with sampling interval $T_s>0$, assuming zero-order hold on the control input $u_i(t)$. The equivalent discrete-time models for \eqref{eq:LTI_C1_Cont} and \eqref{eq:LTI_C2_Cont} can be written compactly as:
\begin{subequations}
\label{eq:LTI_Discrete}
    \begin{align}
        \hat{x}_i(k+1) &= A \hat{x}_i(k) + B_v v_{i-1}(k) + \sum_{j \in \mathcal{L}} \hat{\Gamma}_j \delta_{i,j}(k), \label{eq:LTI_Dis_1} \\
        \bar{x}_i(k+1) &= A \bar{x}_i(k) + B_v v_{i-1}(k) + \sum_{j \in \mathcal{L}} \bar{\Gamma}_j \delta_{i,j}(k) \label{eq:LTI_Dis_2}
    \end{align}
\end{subequations}
with
\begin{equation}
\left\{\begin{aligned}
\label{eq:DiscreteSystemMatrices}
& A=e^{A^c T_s}, B_v = \left(\int_0^{T_s} e^{A^c\left(T_s-s\right)} ds\right) B_v^{c}, \\
& \hat{\Gamma}_j = \left(\int_0^{T_s} e^{A^c\left(T_s-s\right)} ds \right) \hat{\Gamma}_j^c, \, \forall j \in \mathcal{L}, \\
& \bar{\Gamma}_j = \left(\int_0^{T_s} e^{A^c\left(T_s-s\right)} ds \right) \bar{\Gamma}_j^c, \, \forall j \in \mathcal{L}.
\end{aligned}\right.
\end{equation}

\subsection{Adversarial Reachable Sets}
We are primarily interested in resource-limited attacks --- attacks that tamper with sensing, actuation, and networked data while being constrained by factors such as physical limitations, computing power, and attack strategy \cite{zhang_networked_2019}. We model these constraints as hard bounds on attack signals $\delta_{i,j}(k)$. We also impose a hard bound on the velocity of the preceding vehicle, $v_{i-1}$, as it enters the closed-loop dynamics (see \eqref{eq:LTI_Discrete}) as an external disturbance that will affect the system trajectories. Having bounded velocity is reasonable as $v_{i-1}$ is constrained by the physical limitations of the vehicle and highway speed limits. The constraints we impose have the following structure:
\begin{subequations}
\label{eq:Bounds}
    \begin{align}
            & \delta_{i,j}(k) \in \{ \delta_{i,j}(k) \mid \hspace{1mm}  \delta^2_{i,j} \leq W^2_{i,j} \}, \, \forall k \in \mathbb{N}, j \in \mathcal{L},  \label{eq:DeltaBound} \\
            & v_{i-1}(k) \in \{ v_{i-1}(k) \mid \hspace{1mm}  v^2_{i-1} \leq \bar{v}^2\}, \, \forall k \in \mathbb{N},  \label{eq:vBound}
    \end{align}
\end{subequations}
for some known constants $W_{i,j}\in\mathbb{R}_{>0}$ and $\bar{v}\in\mathbb{R}_{>0}$. Associated with these constraints, we introduce the notion of adversarial reachable sets for \eqref{eq:LTI_Dis_1}:
\begin{equation}\label{eq:reachable_set_C1}
\mathcal{R}^{\hat{x}_{i}}(k):=\left\{\hat{x}_{i} \in \mathbb{R}^4  \left|
\begin{aligned}
    & \hat{x}_{i}  = \hat{x}_i(k),  \\ 
    & \hat{x}_i(k) \text{ solution to } \eqref{eq:LTI_Dis_1},\\
    &\delta_{i,j}(k) \text{ satisfies } \eqref{eq:DeltaBound}, \\[1mm]
    & v_{i-1}(k) \text{ satisfies } \eqref{eq:vBound},
\end{aligned}
\right. \right\},
\end{equation}
and for \eqref{eq:LTI_Dis_2}:
\begin{equation}\label{eq:reachable_set_C2}
\mathcal{R}^{\bar{x}_{i}}(k):=\left\{\bar{x}_{i} \in \mathbb{R}^4  \left|
\begin{aligned}
    & \bar{x}_{i}  = \bar{x}_i(k),  \\ 
    & \bar{x}_i(k) \text{ solution to } \eqref{eq:LTI_Dis_2},\\
    &\delta_{i,j}(k) \text{ satisfies } \eqref{eq:DeltaBound}, \\[1mm]
    & v_{i-1}(k) \text{ satisfies } \eqref{eq:vBound},
\end{aligned}
\right. \right\}.
\end{equation}

To quantify the adversarial capabilities of attacks, the volume of these adversarial reachable sets can serve as a security metric \cite{Murguia2020}. This metric provides insight into the size of the state space portion that can be induced by a series of attacks. However, computing the exact value of $\mathcal{R}^{\hat{x}_i}(k)$ is generally not tractable and $k$-dependent. Instead, we seek to obtain the outer ellipsoidal approximation $\mathcal{E}^{\hat{x}_i}(k)$ via Lemma \ref{lemma1}, where due to $a \in (0, 1)$ in Lemma \ref{lemma1}, the sequence $\alpha_i^{\hat{x}_i}(k)$ shaping the ellipsoid $\mathcal{E}^{\hat{x}_i}(k)$ converges to $\alpha_i^{\hat{x}_i}(\infty) \coloneqq (N-a)/(1-a)$ exponentially fast. Therefore, in a few steps, $\mathcal{E}^{\hat{x}_i}(k) \approx \mathcal{E}^{\hat{x}_i}(\infty) \coloneqq \{ \hat{x}_i \mid \hat{x}^\top_i \mathcal{P}^{\hat{x}_i} \hat{x}_i \leq (N-a)/(1-a) \}$, for some positive definite matrix $\mathcal{P}^{\hat{x}_i} \in \mathbb{R}^{4 \times 4}$, and where $N$ is the number of disturbances (according \eqref{eq:Bounds}) acting on the system, hence $ N = 1 + \text{card}[\mathcal{L}]$, where card$[\cdot]$ denotes cardinality. The volume of the ellipsoidal approximation $\mathcal{E}^{\hat{x}_i}(\infty)$ (similarly for $\mathcal{E}^{\bar{x}_i}(\infty)$) is used as an over-approximation of the proposed security metric. Note that, as we consider the reachable set at infinity, the security analysis is independent of the initial condition.

%%%%%%%%%%%%%%%%%%%%%%%%%%%%%%%%%%%%%%%%%%%%%%%%%%%%%%%%%%%%%%%%%%%%%%%%%%%%%%%%
\section{Security Assessment}\label{sec:SecurityAssessment}
%% 8 page version
%\input{V2P8/5.0_SecurityAssessmentP8}

%% 6 page version
In this section, we use Lemma \ref{lemma1} to find the smallest ellipsoidal approximation of the adversarial reachable sets \eqref{eq:reachable_set_C1} and \eqref{eq:reachable_set_C2} for different subsets of sensors \eqref{eq:SensorMeasurements} being attacked. We introduce the notion of critical states, denoted as $\mathcal{D}^{x_i}$, which characterizes states that, if reached, compromise the safety and integrity of the vehicle. These critical states may include situations such as collisions between vehicles or a vehicle exceeding the speed limit on the highway. If the intersection between the set of critical states and a reachable set $\mathcal{R}^{x_i}(k)$ is not empty; then, there exist attack signals $\delta_{i,j}(k)$ satisfying \eqref{eq:DeltaBound} that can drive the vehicle to a critical state. Consider the inter-vehicle distance state $z_i$, and note that $q_{i-1} - q_i - L_i = z_i + r_i \leq 0$ indicates that a collision between vehicles $i$ and $i-1$ has occurred. Therefore, a subset of states $\hat{x}_i$ and $\bar{x}_i$ that represents collision can be written as $z_i \leq r_i$. Similarly, the vehicle velocities $v_i$ exceeding a speed limit, say $\bar{v}_i$, can be formulated as $\tfrac{1}{h}z_i - \tfrac{1}{h}e_i > \bar{v}_i$. 

These two sets of critical states and ellipsoidal approximations, projected onto the $v_i$-$z_i$ plane via \text{Lemma \ref{lemma2}} and applying a coordinate transformation $\tilde{x} = \begin{bmatrix}
    v_i & z_i & \dot{e}_i & \zeta_{i,1}
\end{bmatrix}^\top = S \hat{x}$, (similarly for $\tilde{x}$), such that, $\mathcal{E}^{\tilde{x}_i}(\infty) \coloneqq \{ \tilde{x}_i \!\! \mid \!\! \tilde{x}^\top_i (S^{-1})^\top \mathcal{P}^{\hat{x}_i} S^{-1} \tilde{x}_i \leq (N-a)/(1-a) \}$. This transformation allows the characterizing of security using the size of the ellipsoids and their intersection with critical states. Note that the proposed critical states are used as an example, but the analysis can also be applied to different safety constraints.

 We consider the closed-loop dynamics \eqref{eq:LTI_Dis_1}, \eqref{eq:DiscreteSystemMatrices} for $\mathcal{C}_1$, and \eqref{eq:LTI_Dis_2}, \eqref{eq:DiscreteSystemMatrices} for $\mathcal{C}_2$, with a desired inter-vehicle distance of $r_i = 3\,$m, driveline dynamics constant $\tau = 0.1\,$s, time headway constant $h = 0.5\,$s, controller gains of $(k_p, k_d, k_{dd}) = (0.2, 0.7, 0)$, and sampling rate of $T_s = 0.01\,$s. Additionally, we assume a speed limit of $\bar{v}_i = 35.83 \,$m/s. For the resource-limited FDI attacks $\delta_{i,j}(k)$, we assume that the attacks remain within the bound specified by \eqref{eq:DeltaBound} with $W_{i,j} = 1$, for all $j\in \mathcal{L}$. The bound $W_{i,j}$ is an arbitrarily chosen value, as we are only interested in the differences in the proposed metrics between the two controller realizations.

%% 8 page version
%\input{V2P8/5.1_SensorSensitivityP8}

%% 6 page version
\subsection{Sensor Sensitivity}
\label{sec:SensorSensitivity}
\begin{figure}[bt]\centering
		\includegraphics[width=0.9\linewidth]{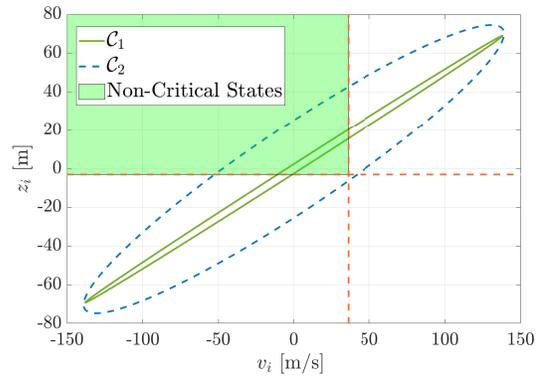}
		\caption{Projection of $\mathcal{E}^{\hat{x}_i}(\infty)$ and $\mathcal{E}^{\bar{x}_i}(\infty)$ for an attack on acceleration sensor $y_{i,3}$, hence $\mathcal{L} = 3$.}
		\centering
		\label{fig:Att_y3}
    \vspace{-3.9mm}
\end{figure}

The first case study focuses on individual sensor attacks, offering valuable insights into the importance of protecting specific sensors during cyber attacks. This case study also indicates which sensors are more critical to be protected when there are resource limitations and not all sensors can be secured. In Table \ref{tab:Results1}, the numerical values of volumes of $\mathcal{E}^{\hat{x}_1}(\infty)$ and $\mathcal{E}^{\bar{x}_1}(\infty)$ (i.e., for both realized controllers $\mathcal{C}_1$ and $\mathcal{C}_2$) are given. Upon comparing the different individual attacks, $\mathcal{C}_2$ proves to be highly sensitive to an attack on the onboard acceleration measurement $y_{i,3}$. As for the realization it holds that $k_{dd}=0$, resulting in an equivalent attack on $\xi_i$ when comparing $\hat{\Gamma}_{3}^c$ and $\bar{\Gamma}_{3}^c$. However, due to the different realization, the error dynamics $\ddot{e}$ in $\mathcal{C}_2$ is also affected by an attack on $y_{i,3}$, hence the non-zero entry in $\bar{\Gamma}_{3}^c$, resulting in an increased sensitivity to an attack on this particular sensor. In Fig. \ref{fig:Att_y3}, the projection of $\mathcal{E}^{\hat{x}_1}$ and $\mathcal{E}^{\bar{x}_1}$ (onto the $v_i$-$z_i$ plane) for $\mathcal{L} = 3$ is shown. The orange dashed lines indicate the boundaries of the critical states, where it can be observed that $\mathcal{C}_2$ suffers significantly more as the number of states the attack can induce is much larger. 

Based on the comparison of the different volumes in \text{Table \ref{tab:Results1}}, it can be concluded that $\mathcal{C}_1$ has greater overall robustness against resource-limited attacks. The attack in which $\mathcal{C}_2$ outperforms $\mathcal{C}_1$ is in the case of a cyber-attack on $u_{i-1}$ ($\mathcal{L} = 6$). However, it should be noted that an attack on $u_{i-1}$ is equivalent to an attack on $a_{i-1}$, as both values are typically obtained through V2V communication. As a result, the comparison between scenarios $\mathcal{L} = 5$ and $\mathcal{L} = 6$ does not indicate any advantage of using one realization over the other. Furthermore, it is worth noting that for $\mathcal{C}_1$, the information of the predecessor --- namely $v_{i-1}$ and $u_{i-1}$ --- is important to protect, especially when resources are limited. This emphasizes the need for enhanced security against cyber-attacks, highlighting the importance of safeguarding such critical information. 

\begin{table}[hbt]
\centering
\begin{tabular}{l|l|l|l|l|l|l|}
$\mathcal{L}$ & 1 & 2 & 3 & 4 & 5 & 6 \\ \hline
Vol($\mathcal{E}^{\hat{x}_i}$)   & 192.92  & 96.46 & 337.64 & 675.59 & 0.01 & 965.73 \\
Vol($\mathcal{E}^{\bar{x}_i}$)   & 192.92  & 96.46 & 3523.42 & 675.59 & 951.81 & 0.01  
\end{tabular}
    \caption{Numerical values of volumes of $\mathcal{E}^{\hat{x}_i}(\infty)$ and $\mathcal{E}^{\bar{x}_i}(\infty)$ (i.e., for both realized controllers $\mathcal{C}_1$ and $\mathcal{C}_2$).}\label{tab:Results1}
    \vspace{-8mm}
\end{table}

%% 8 page version
%\input{V2P8/5.2_ConstantHeadwaySensitivityP8}

%% 6 page version
\subsection{Varying Constant Headway $h$}
In this section, we present a second case study, which explores the impact of the time headway constant $h$ on the ellipsoidal approximation of the adversarial reachable set for the same system introduced in section \ref{sec:SensorSensitivity}. 

\begin{figure}[bt]\centering
		\includegraphics[width=1.1\linewidth]{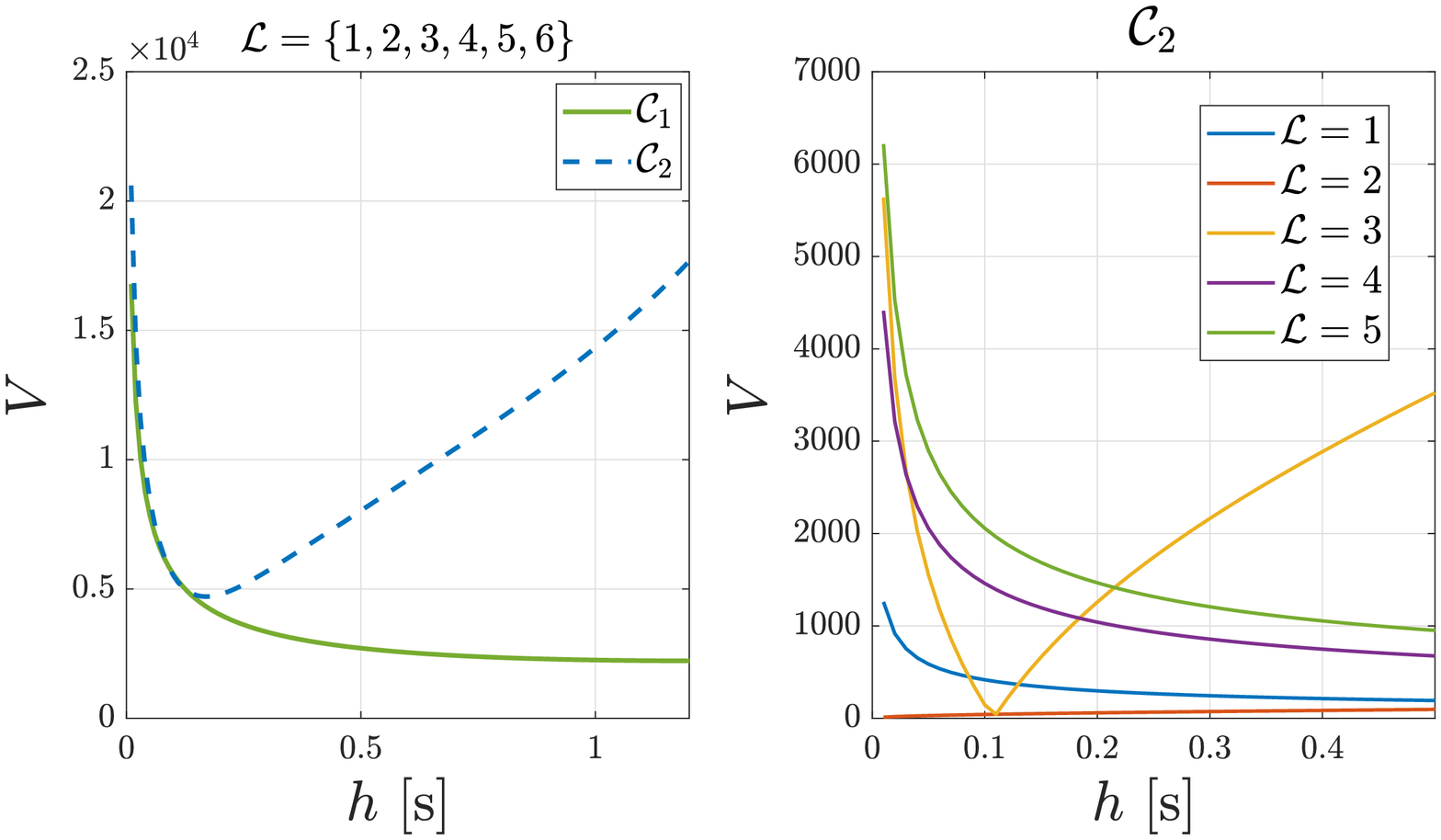}
		\caption{(Left) The volume of $\mathcal{E}^{\hat{x}_i}(\infty)$ and $\mathcal{E}^{\bar{x}_i}(\infty)$ for $\mathcal{L} = \{1,2,3,4,5,6\}$ for varying headway constant $h$. (Right) The volume of $\mathcal{E}^{\bar{x}_i}(\infty)$ for varying headway constant $h$ and individual sensor attacks.}
		\centering
		\label{fig:VolumeC1C2_hSweep}
  \vspace{-6.75mm}
\end{figure}

We assume that $\mathcal{L} = \{1,2,3,4,5,6\}$, indicating that all sensors in \eqref{eq:SensorMeasurements} are compromised. The left plot in \text{Fig. $\,$\ref{fig:VolumeC1C2_hSweep}} shows the projected ellipsoids' volume of both $\mathcal{E}^{\hat{x}_i}(\infty)$ and $\mathcal{E}^{\bar{x}_i}(\infty)$ on the $v_i$-$z_i$ plane for different values of $h \in [0.01, 0.02,..., 1.2]$. The results show that the volume of $\mathcal{E}^{\hat{x}_i}$ has an exponential decay for $h < 1$, indicating more resiliency, however for $h>1$ the volume slightly increases. It is observed that the volume of $\mathcal{E}^{\hat{x}_i}$ is always smaller than $\mathcal{E}^{\bar{x}_i}$, but decays more slowly. However, for $h>0.1$, the volume of $\mathcal{E}^{\bar{x}_i}$ increases exponentially. To analyze this behavior, the right plot in Fig. \ref{fig:VolumeC1C2_hSweep} shows the volume of  $\mathcal{E}^{\bar{x}_i}$ for single sensor attacks. It is clear that for $\mathcal{L} = 3$ and $h>0.1$, the volume begins to increase exponentially. Since $k_{dd}$ is set to be zero, the only difference between $\hat{\Gamma}_3$ and $\bar{\Gamma}_3$ is the additional term $1-\frac{h}{\tau}$, where the volume starts increasing when this term switches to a negative term. Future work should include an analysis of why the additional term in $\bar{\Gamma}_3$ causes exponential growth of the volume.

Additionally, it is noted that the volume alone does not provide a representative metric for selecting the optimal value of $h$. In Fig. \ref{fig:VolumeC1C2_hSweep} $\mathcal{E}^{\hat{x}_i}(\infty)$ and $\mathcal{E}^{\bar{x}_i}(\infty)$ are projected onto the $v_i$-$z_i$ plane, showing that the orientation of the ellipsoids are affected for different values of $h$. Thus, the optimal controller does not necessarily have the smallest volume, but has the smallest intersection with the critical states, indicating a potential new security metric.

\begin{figure}[bt]\centering
		\includegraphics[width=1.1\linewidth]{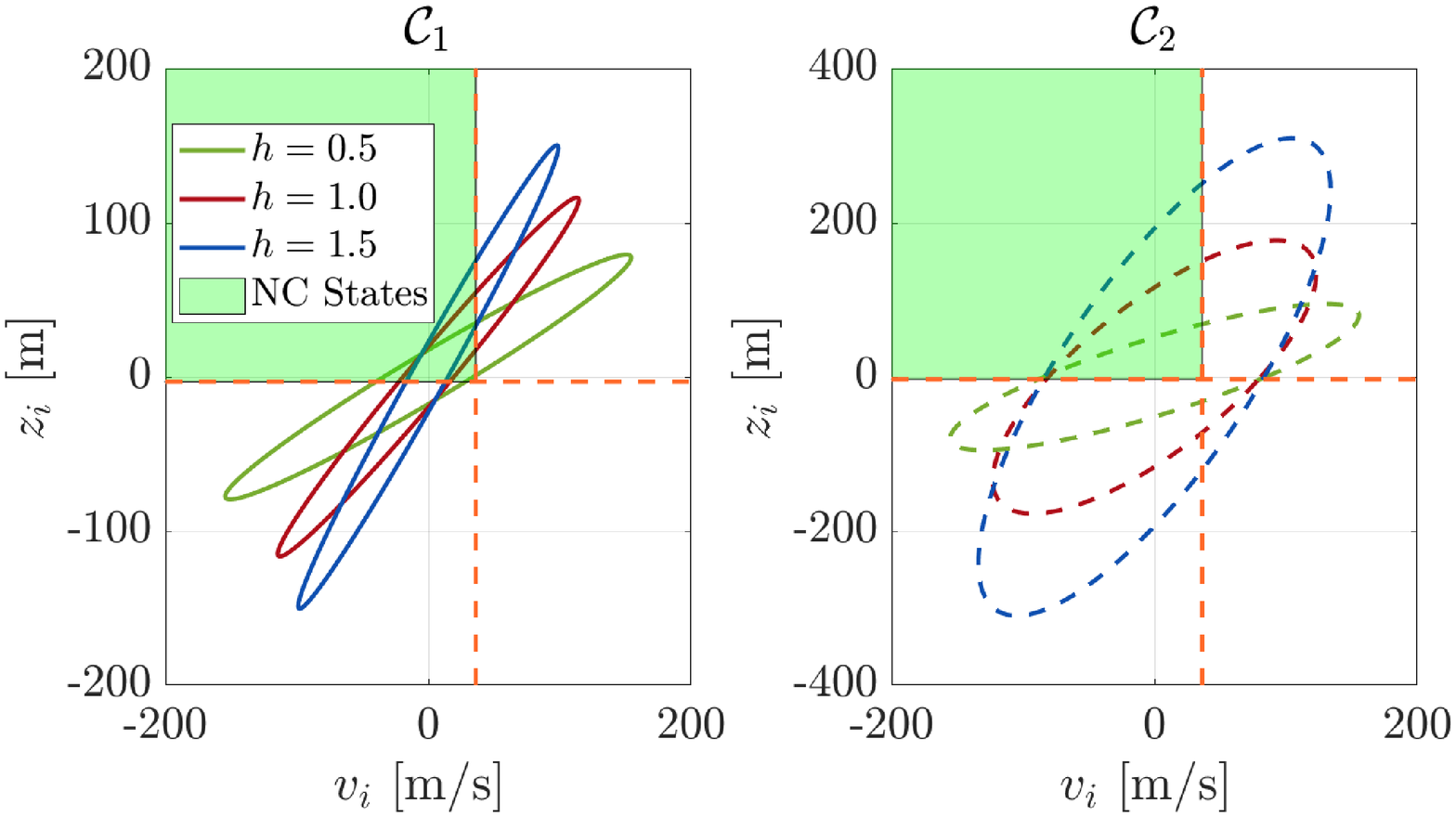}
		\caption{(Left) Projection $\mathcal{E}^{\hat{x}_i}(\infty)$ onto $v_i$-$z_i$ plane for different $h$ and $\mathcal{L} = \{1, 2, 3, 4, 5, 6\}$. (Right) Projection $\mathcal{E}^{\bar{x}_i}(\infty)$ onto $v_i$-$z_i$ plane for different $h$ and $\mathcal{L} = \{1, 2, 3, 4, 5, 6\}$. The green area represents the Non-Critical States (NC-States).}
		\centering
		\label{fig:Ellipses_C1C2_hSweep}
    \vspace{-3mm}
\end{figure}

%% 8 page version
%\input{V2P8/5.3_SamplingSensitivityP8}

%% 6 page version
\subsection{Varying Sampling Constant $T_s$}
This section discusses the final case study, where the effect of the sampling rate $T_s$ on $\mathcal{E}^{\hat{x}_i}(\infty)$ and $\mathcal{E}^{\bar{x}_i}(\infty)$ is examined. Since the CACC can be applied to various vehicles / operating systems, it is most likely for the sampling rate to differ. Assuming a zero-order hold for the controller input, the closed-loop system performance can be affected. 

\begin{figure}[bt]\centering
		\includegraphics[width=1.1\linewidth]{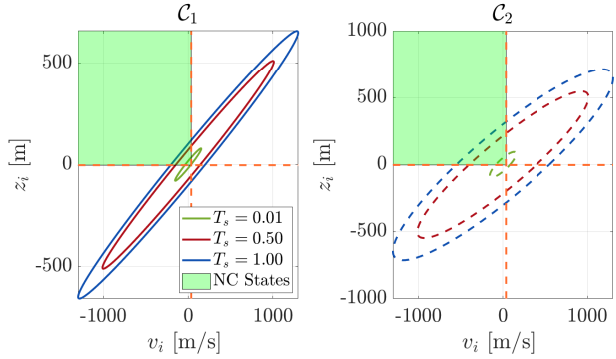}
		\caption{(Left) Projection $\mathcal{E}^{\hat{x}_i}(\infty)$ onto $v_i$-$z_i$ plane for different $T_s$ and $\mathcal{L} = \{1, 2, 3, 4, 5, 6\}$. (Right) Projection $\mathcal{E}^{\bar{x}_i}(\infty)$ onto $v_i$-$z_i$ plane for different $T_s$ and $\mathcal{L} = \{1, 2, 3, 4, 5, 6\}$. The green area represents the Non-Critical States (NC-States).}
		\centering
		\label{fig:Ellipse_C1C2_TsSweep}
    \vspace{-6.5mm}
\end{figure}

The obtained results are similar to the previous two case studies, where in Fig. \ref{fig:Ellipse_C1C2_TsSweep} $\mathcal{E}^{\hat{x}_i}(\infty)$ and $\mathcal{E}^{\bar{x}_i}(\infty)$ are projected onto the $v_i$-$z_i$ plane for different values of $T_s$, indicating that $\mathcal{C}_1$ is more resilient against attacks than $\mathcal{C}_2$ for varying $T_s$. However, where $h$ also affected the orientation of the ellipsoid, $T_s$ only affects the size of the ellipsoid. Since $T_s$ only alters the input rate, it is expected to only affect the volume, as changing $h$ results in a different control goal and increased distance between vehicle $i$ and its predecessor $i-1$.

%%%%%%%%%%%%%%%%%%%%%%%%%%%%%%%%%%%
%\begin{figure}[bt]\centering
%		\includegraphics[width=1.1\linewidth]{V2P6/Figures/figure_TsSweep.eps}
%		\caption{The volume of the ellipsoidal approximation $\mathcal{E}^{\hat{x}_i}(\infty)$ and $\mathcal{E}^{\bar{x}_i}(\infty)$ for $\mathcal{L} = \{1,2,3,4,5,6\}$ for varying sampling constant $T_s$.}
%		\centering
%		\label{fig:TsSweep}
%    \vspace{-3mm}
%\end{figure}

%%%%%%%%%%%%%%%%%%%%%%%%%%%%%%%%%%%%%%%%%%%%%%%%%%%%%%%%%%%%%%%%%%%%%%%%%%%%%%%%
\section{CONCLUSIONS AND FUTURE WORKS}\label{sec:Conclusion}
%% 8 page version
%\input{V2P8/6.ConsclusionFutureworkP8}

%% 6 page version
CACC schemes must ensure safety and reliability in adversarial environments. In this paper, we have argued that for a given dynamic CACC scheme, an infinite number of real-time realizations of the same controller exist given the available sensors. It is shown that different controller realizations can significantly affect the impact of resource-limited attacks. Two different controller realizations are compared by computing outer ellipsoidal approximations of reachable sets induced by attacks, and evaluated how the sampling and headway constant change these sets. As a result, it is concluded that the sensitivity of the system changes significantly for the same class of attacks. Therefore, these results indicate that there exist optimal controllers that minimize the system sensitivity to resource-limited attacks.

In this manuscript, we have also highlighted the importance of incorporating new security metrics. Changing the controller affects the orientation of the ellipsoidal approximation, resulting in different intersections between critical states and adversarial reachable sets. Therefore, comparing different controller realizations by means of these intersections provides more insight into the vulnerabilities of the system in terms of safety. Additionally, future work should incorporate an analysis of how an attack propagates through the platooning behavior by considering more vehicles.

%%%%%%%%%%%%%%%%%%%%%%%%%%%%%%%%%%%%%%%%%%%%%%%%%%%%%%%%%%%%%%%%%%%%%%%%%%%%%%%%
%\section{ACKNOWLEDGMENTS}

%Thanks

%%%%%%%%%%%%%%%%%%%%%%%%%%%%%%%%%%%%%%%%%%%%%%%%%%%%%%%%%%%%%%%%%%%%%%%%%%%%%%%%

\end{document}